# Understanding the relationship between visual-angle and eye-angle for reliable determination of the field-of-view in ultra-wide field fundus photography

**XINCHENG YAO[1,2,\*], DEVRIM TOSLAK[3], TAEYOON SON[1] AND JIECHAO MA[1]**

[1]*Department of Bioengineering, University of Illinois at Chicago, Chicago, IL 60607, USA*
[2]*Department of Ophthalmology and Visual Sciences, University of Illinois at Chicago, Chicago, IL 60612, USA*
[3]*Antalya Training and Research Hospital, Department of Ophthalmology, Antalya, 07100, Turkey*
*\*xcy@uic.edu*

**Abstract:** Visual-angle has been used as the conventional unit to determine the field-of-view (FOV) in traditional fundus photography. Recently emerging usage of eye-angle as the unit in wide field fundus photography creates confusions about FOV interpretation in instrumentation design and clinical application. This study is to systematically derive the relationship between the visual-angle $\theta_v$ and eye-angle $\theta_e$, and thus to enable reliable determination of the FOV in wide field fundus photography. FOV conversion ratio $\theta_e/\theta_v$, angular conversion ratio $\Delta\theta_e/\Delta\theta_v$, retinal conversion ratio $\Delta d/\Delta\theta_v$, retinal distance and area are quantitatively evaluated. Systematical analysis indicates that reliable conversion between the $\theta_v$ and $\theta_e$ requires determined nodal point and spherical radius of the eye; and the conversion ratio is not linear from the central field to peripheral region. Based on the eye model with average parameters, both angular conversion ($\Delta\theta_e/\Delta\theta_v$) and retinal conversion ($\Delta d/\Delta\theta_v$) ratios are observed to have a 1.51-fold difference at the central field and far peripheral region. A conversion table, including $\theta_e/\theta_v$, $\Delta\theta_e/\Delta\theta_v$, $\Delta d/\Delta\theta_v$, retinal area and percentage ratio, is created for reliable assessment of imaging systems with variable FOV.

## 1. Introduction

Fundus photography is important for eye disease screening, diagnosis and treatment assessment. In traditional fundus cameras, visual-angle $\theta_v$ is used for quantitative evaluation of the field-of-view (FOV). For visual perception, the visual-angle is the angle of a target subtends at the eye. Under emmetropic condition, the normal eye can image distant objects precisely in focus into the retina [1]. At the nodal point of the eye, the angle subtended by the target image at the retina is identical to the visual-angle $\theta_v$ (Fig. 1(a)). In other word, the incident light from each direction corresponds to a focused point at the retina. For fundus photography, the purpose of retinal imager is to collect light from the back of the eye. According to the principle of reversibility of light, the light coming from each retinal point corresponds to a specific direction of the light leaving the eye (Fig. 1(b)). The light angle range detectable by the fundus imager defines the FOV. Therefore, the FOV and also angular resolution of a fundus imager can be readily calibrated in the unit of visual-angle $\theta_v$, by imaging a target placed at a certain distance [2].

The FOV of traditional fundus cameras is typically limited at 30° to 60° visual-angle [3]. In principle, eye diseases may affect any part of the retina. The retina covers 72% of the inside of the eye globe [4], or roughly 180° visual field. In order to achieve the necessary view field coverage, mydriatic ETDRS 7-field photography for screening diabetic retinopathy (DR) has been developed based on the use of the traditional fundus camera with a 30° visual-angle FOV [5]. Conventional binocular indirect ophthalmoscopy (BIO), which is a time-consuming procedure that is painful for the patient and stressful for the ophthalmologist, is still the gold standard for reliable screening of retinopathy of prematurity (ROP) screening [6-8]. In order to foster digital fundus photography for eye disease detection and treatment management, there are active efforts to develop wide field imaging system [9-17]. Instead of using traditional visual-angle to specify the FOV, some of these wide field systems such as the Optos (Optos, Marlborough, MA), Retcam (Clarity Medical Systems, Pleasanton, CA) and Clarus 500 (Carl Zeiss Meditec, Inc., Dublin, CA) systems use eye-angle $\theta_e$ to specify the FOV. As shown in Fig. 1(a), the eye-angle $\theta_e$ is the angle subtended by the imaged retinal region at the spherical center of the eye.

Given the factor of that not all researchers, ophthalmologists and photographers are aware of the difference between visual-angle and eye-angle, the mixed usages of these two terms have produced confusions about the visual field interpretation. In principle, the visual-angle $\theta_v$ and eye-angle $\theta_e$ can be converted to each other. One empirical factor ~1.5 has been used to convert the visual and eye angles as $\theta_e=1.5\theta_v$ [13-16, 18-20]. The empirical factor ~1.5 was estimated based on the refractive difference between the air and the eye which is assumed with 11 mm spherical radius [18]. Without considering detailed properties, such as optical powers of the cornea and crystalline lens, the simplified model in the white paper [18] is not sufficient to reliably correlate the visual-angle and eye-angle values over the visual field. In other words, the empirical factor ~1.5 is reasonable only for a certain angle range nearby the eye axis, and can produce overestimated eye-angle number when the visual-angle increases to the peripheral retina. In this article, we report a systematic analysis of the relationship between visual-angle and eye-angle from the optical axis to the ora serrata, i.e. the far peripheral region of the retina.

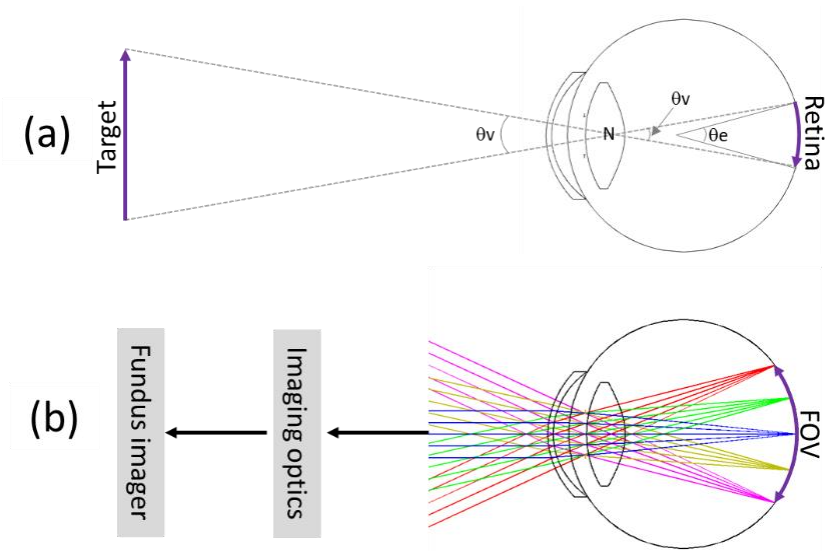

Fig. 1. (a) Illustration of visual-angle and eye-angle. (b) Illustration of FOV.

## 2. Materials and Methods

Figure 2(a) illustrates the eye model used for following analysis. The parameters in Fig. 2(a) have been popularly reported in previous publications [1, 21, 22], although slight difference may exist among different studies. Because every eye is intrinsically unique, these parameters represent average values for the purpose of generalization [1]. As shown in Fig. 2(a), the cornea is the mostly outer layer of the eye. The cornea is a spherical section with an anterior radius of curvature of 7.8 mm, posterior radius of curvature of 6.7 mm, thickness of 0.55 mm, and refractive index of 1.3771. The crystalline lens is a biconvex lens with 10 mm and -6 mm radii of curvature for the anterior and posterior surfaces. The refractive indices of the aqueous, crystalline lens, and vitreous humors are 1.3374, 1.42 and 1.336, respectively. The distance between the posterior surface of the cornea to the anterior surface to the crystalline lens is 3.1 mm, and the thickness of the crystalline lens is 4.0 mm. The radius of posterior eye is 11 mm.

The nodal point is the location in the eye where light entering or leaving the eye and passing through is undeviated. This allows the same visual-angle to be used to determine the retinal region corresponding to the target in external space (Fig. 1). The nodal point of the eye is known to be nearby the posterior surface of the crystalline lens, i.e., anterior vitreous [23, 24] (Fig. 2(a)). Using the parameters in Fig. 2(a), a ZEMAX simulation confirms the lens posterior as the nodal point in the eye (Fig. 2(b)). Fig. 2(c) is used to explain the relationship between visual-angle $\theta_v$ and eye-angle $\theta_e$. In order to simplify the discussion, we only use the space above the optical axis to illustrate the half visual-angel $\theta_v'$ and half eye-angle $\theta_e'$. The same relationship can be applied to the bottom half of the space. In other words, the FOV can be simply doubled compared to the half space values in Fig. 2(c), i.e., $\theta_v = 2\theta_v'$ and $\theta_e = 2\theta_e'$.

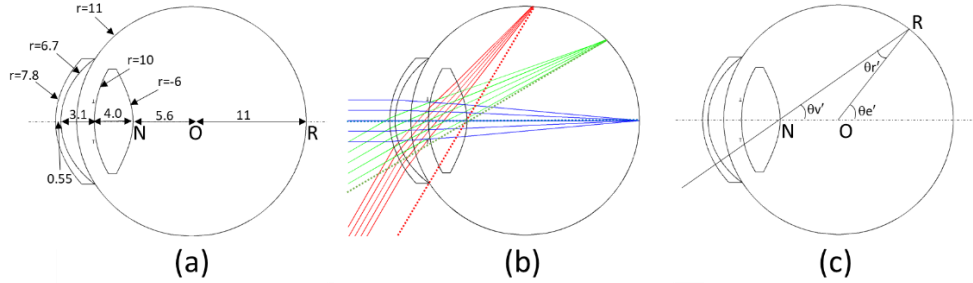

Fig. 2. (a) Schematic diagram of the eye model. (b) Ray tracing through the eye model. (c) Illustration of the visual-angel and eye-angle.

As shown in Fig. 2(c), the half eye-angle $\theta_e'$ can be estimated as follows:

$$\theta_e' = \theta_v' + \theta_r' \tag{1}$$

where $\theta_v'$ is the half visual-angle, and

$$\theta_r' = \sin^{-1}(\frac{NO}{OR} \sin \theta_v') \tag{2}$$

Substitute NO=5.6 mm and OR=11 mm into Equation 2

$$\theta_r' = \sin^{-1}(0.51 \sin \theta_v') \tag{3}$$

Substitute Equation 3 into Equation 1

$$\theta_e' = \theta_v' + \sin^{-1}(0.51 \sin \theta_v') \tag{4}$$

If the FOV is known in the unit of eye-angle, corresponding retinal distance can be estimated as a spherical arc (Fig. 3).

$$d = 2\pi r \theta_e'/360 = \pi r \theta_e'/180 \tag{5}$$

where r is the eyeball radius.

Substitute Equation 4 into 5,

$$d = \pi r (\theta_v' + \sin^{-1}(0.51 \sin \theta_v'))/180 \tag{6}$$

Corresponding retinal area A can be estimated as a spherical cap (Fig. 3).

$$A = 2\pi r h \tag{7}$$

and h is the height of the spherical cap corresponding to the visual field. The h can be calculated as

$$h = (1 - \cos \theta_e')r \tag{8}$$

Substitute Equation 6 into Equation 5

$$A = 2\pi r^2 (1 - \cos \theta_e') \tag{9}$$

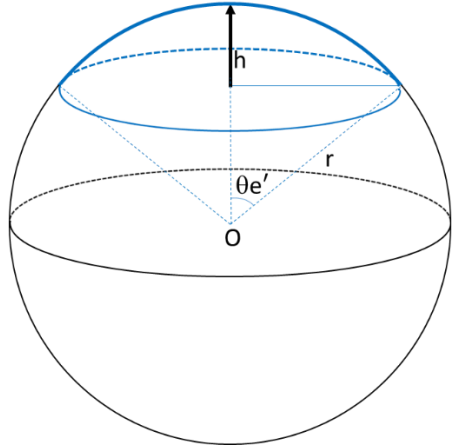

Fig. 3. Illustration of retinal area estimation as a spherical cap.

## 3. Results

According to Equation 4, Fig. 4(a) show the relationship between visual-angel $\theta_v$ and eye-angle $\theta_e$. Moreover, Fig. 4(b) illustrate the FOV conversion ratio $\theta_e/\theta_v$ and angular conversion ratio $\Delta\theta_e/\Delta\theta_v$ from the central to peripheral regions. As shown in Fig. 4(a), the relationship between visual-angle and eye-angle FOVs is not linear. Figure 4(b) confirm the FOV conversion ratio $\theta_e/\theta_v$ is ~1.5 for FOV values within 75º, but the factor gradually decreases to 1.34 for large angle FOV. The 180º visual-angle FOV, which is roughly the total visual field, corresponds to 241º eye-angle FOV (Fig. 4(a)).

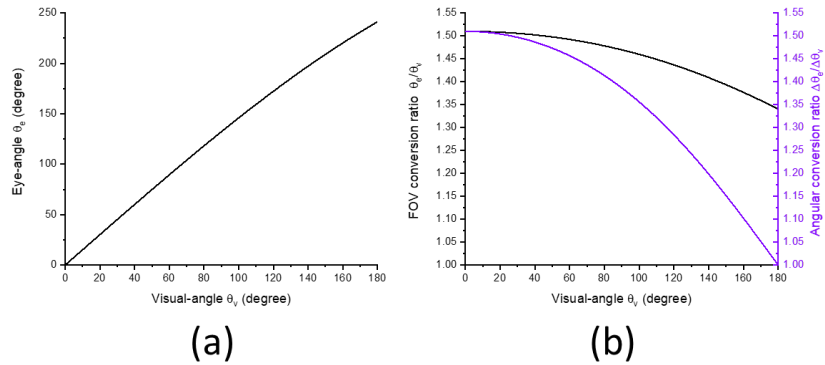

Fig. 4. (a) The relationship between visual-angel $\theta_v$ and eye-angle $\theta_e$. (b) Conversion ratio $\theta_e/\theta_v$. (c) $\Delta\theta_e/\Delta\theta_v$ at different visual field.

In order to further evaluate the angular conversion factor $\Delta\theta_e/\Delta\theta_v$, the first order derivation of the Equation 4 is also implemented. As shown in Fig. 4(b), the angular conversion factor

$\Delta\theta_e/\Delta\theta_v$ is ~1.51 at the central field, i.e., nearby the optical axis of the eye. The conversion factor can be reduced up to 1.0 at the far periphery, i.e. the ora serrata.

Equation 5 reveals that the imaged retinal distance d is linearly proportional to the eye-angle value. Assume r=11 mm, the retinal conversion ratio $\Delta d/\Delta\theta_e$ can be estimated at 192 μm per eye-angle degree, which is constant over the whole visual field. In contrary, Equation 6 reveals that the imaged retinal distance d is not linearly proportional to the eye-angle value. According Equation 6, Fig. 5(a) show the relationship between the retinal distance d and visual-angle $\theta_v$. At the center of the visual field, by considering the angular conversion factor $\Delta\theta_e/\Delta\theta_v$ 1.51 (Fig. 4), 290 μm per visual-angle degree can be estimated. The 290 μm per visual-angle degree is reasonably consistent to the popularly known 288 μm per visual-angle degree in previous publications [4]. However, the angular conversion factor $\Delta\theta_e/\Delta\theta_v$ is gradually reduced, up to 192 μm/degree at the far peripheral visual field (Fig. 5b).

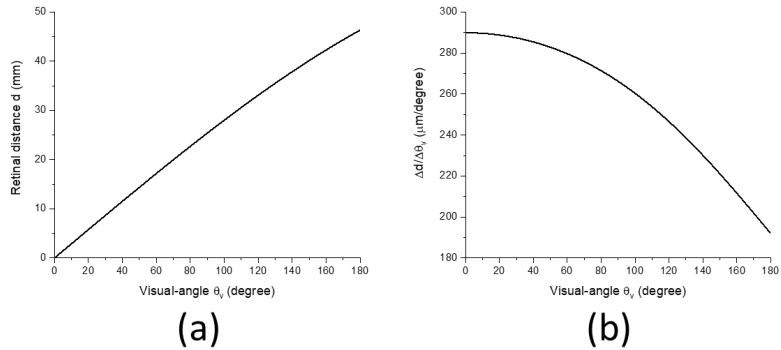

Fig. 5. (a) Relationship between retinal distance d and visual-angle $\theta_v$. (b) $\Delta d/\Delta\theta_v$ at different visual field.

Figure 6 shows the relationships of visual-angel FOV (Fig. 6(a)) and eye-angle FOV (Fig. 6(b)) to retinal area and percentage ratio. As shown in Fig. 6(a), the 180° visual-angle, i.e., 241° eye-angle, corresponds to 1,148 mm$^2$, which is reasonably close to previously reported retinal area 1,094 mm$^2$ [4] and 1,134 mm$^2$ [25].

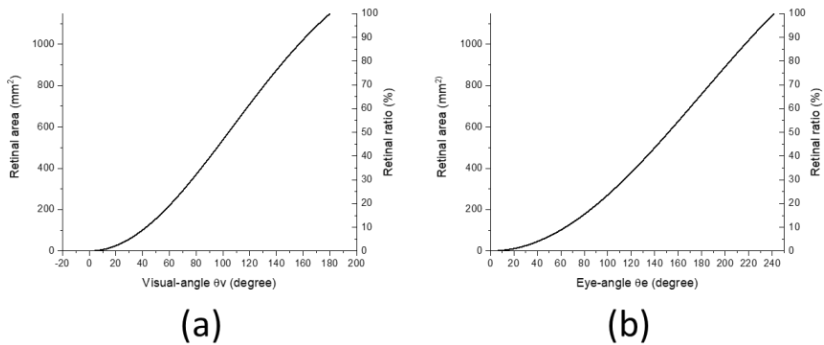

Fig. 6. (a) Relationship between retinal area and visual-angle FOV. (b) Relationship between retinal area and eye-angle FOV.

For easy reference, Table 1 shows the representative values of visual-angle, eye-angle, conversion factor and retinal area, corresponding to variable FOVs. 30° and 45° visual-angle FOVs are mostly popular for traditional fundus cameras. As shown in Table 1, the 30° visual-angle corresponds to a 45.2º eye-angle; and the 45º visual-angle corresponds to a 67.5º eye-angle. A 145º visual-angle corresponds to a 203º eye-angle FOV.

Table 1. Conversion ratios of visual-angle, eye-angle, retinal dimension corresponding to visual field

| $\theta_v(º)$ | $\theta_e(º)$ | $\theta_e/\theta_v$ | $\Delta\theta_e/\Delta\theta_v$ | $\Delta d/\Delta\theta_v$ ($\mu m$/degree) | Retinal area ($mm^2$) | Retinal ratio (%) |
|---|---|---|---|---|---|---|
| 0 | 0 | 1.51 | 1.51 | 290 | 0 | 0 |
| 5 | 7.5 | 1.51 | 1.51 | 290 | 1.65 | 0.14 |
| 10 | 15.1 | 1.51 | 1.51 | 290 | 6.7 | 0.57 |
| 15 | 22.6 | 1.51 | 1.51 | 289 | 14.8 | 1.3 |
| 20 | 30.2 | 1.51 | 1.51 | 289 | 26.2 | 2.3 |
| 25 | 37.7 | 1.51 | 1.50 | 288 | 40.7 | 3.5 |
| 30 | 45.2 | 1.51 | 1.50 | 287 | 58.3 | 5.1 |
| 35 | 52.6 | 1.50 | 1.49 | 286 | 78.8 | 6.9 |
| 40 | 60.1 | 1.50 | 1.49 | 285 | 102.2 | 8.9 |
| 45 | 67.5 | 1.50 | 1.50 | 284 | 128.2 | 11.2 |
| 50 | 74.9 | 1.50 | 1.47 | 283 | 156.7 | 13.6 |
| 55 | 82.2 | 1.50 | 1.47 | 281 | 187.6 | 16.3 |
| 60 | 89.5 | 1.49 | 1.48 | 280 | 220.1 | 19.2 |
| 65 | 96.8 | 1.49 | 1.45 | 278 | 255.5 | 22.3 |
| 70 | 104.0 | 1.49 | 1.44 | 276 | 292.3 | 25.5 |
| 75 | 111.2 | 1.48 | 1.43 | 274 | 330.6 | 28.8 |
| 80 | 118.3 | 1.48 | 1.41 | 271 | 370.3 | 32.3 |
| 85 | 125.3 | 1.47 | 1.40 | 269 | 411.0 | 35.8 |
| 90 | 132.3 | 1.47 | 1.39 | 266 | 452.7 | 39.4 |
| 95 | 139.2 | 1.47 | 1.37 | 263 | 495.1 | 43.1 |
| 100 | 146.0 | 1.46 | 1.36 | 260 | 538.0 | 46.9 |
| 105 | 152.7 | 1.45 | 1.34 | 257 | 581.1 | 50.6 |
| 110 | 159.4 | 1.45 | 1.32 | 254 | 624.2 | 54.4 |
| 115 | 166.0 | 1.44 | 1.30 | 250 | 667.3 | 58.1 |
| 120 | 172.4 | 1.44 | 1.28 | 247 | 710.0 | 61.8 |
| 125 | 178.8 | 1.43 | 1.26 | 243 | 752.3 | 65.5 |
| 130 | 185.1 | 1.42 | 1.24 | 239 | 793.8 | 69.1 |
| 135 | 191.2 | 1.42 | 1.22 | 234 | 834.6 | 72.7 |
| 140 | 197.3 | 1.41 | 1.20 | 230 | 874.4 | 76.2 |
| 145 | 203.2 | 1.40 | 1.18 | 226 | 913.2 | 79.5 |
| 150 | 209.0 | 1.39 | 1.15 | 221 | 950.8 | 82.8 |
| 155 | 214.7 | 1.39 | 1.13 | 216 | 987.1 | 86.0 |
| 160 | 220.3 | 1.38 | 1.10 | 212 | 1022.1 | 89.0 |
| 165 | 225.7 | 1.37 | 1.08 | 207 | 1055.8 | 92.0 |
| 170 | 231.1 | 1.36 | 1.05 | 202 | 1088.0 | 94.8 |
| 175 | 236.3 | 1.35 | 1.03 | 197 | 1118.8 | 97.5 |
| 180 | 241.3 | 1.34 | 1.00 | 192 | 1148.0 | 100 |

## 4. Discussion

In summary, the relationship between visual-angle $\theta_v$ and eye-angle $\theta_e$ has been derived based on the eye model. As shown in Fig. 2(c) and Equations 1-4, the conversion between visual-angle $\theta_v$ and eye-angle $\theta_e$ dependent on the distance between the nodal point N to spherical center O and the radius r of the eye. With the average values, i.e., 5.6 mm from N to O and 11 mm radius, the relationship between visual-angle $\theta_v$ and eye-angle $\theta_e$ has quantitatively evaluated. The relationship between visual-angle and eye-angle is not linear (Fig.4). The FOV conversion factor $\theta_v/\theta_e$ is changed from 1.51 at the central field, i.e., nearby the optical axis, to 1.34 at the far peripheral field, i.e. ora serratta. Similarly, the angular conversion factor $\Delta\theta_v/\Delta\theta_e$ is changed from 1.51 to 1.0 at the far peripheral field.

The empirical factor FOV conversion factor ~1.5 has been used to convert visual-angle $\theta_v$ into eye-angle $\theta_e$ in previous publications [13-16, 18, 19]. As shown in Table 1 and Fig. 4, the FOV conversion factor 1.5 is reasonable for visual-angle $\theta_v<80°$. Therefore, for traditional fundus cameras with typical FOV of 30° or 45° visual-angle, the conversion $\theta_e=1.5\theta_v$ is appropriate. The 30° and 45° visual-angles correspond to 45.2° and 67.5° eye-angles, respectively. The determined FOV in the unit of eye-angle can be readily used to estimate corresponding retinal distance and area. Individual snapshot 30° and 45° visual-angle FOVs correspond to 5.1% and 11.2% of the whole retinal area, respectively. For the wide field imaging systems with 60-80° visual-angle FOV [15, 16, 20], it is also acceptable to use the empirical factor ~1.5 to estimate eye-angle FOV.

However, for ultra-wide field imaging systems [13, 14], direct use of the empirical conversion factor 1.5 might produce overestimated FOV in the unit of eye-angle, and thus overestimated retinal area. For example, the 180° visual-angle FOV corresponds to 241.3° in Table 1, with conversion factor 1.34 derived symmetrically. The 241° eye-angle FOV corresponds to the whole retinal area of 1,148 mm$^2$, which is reasonably close to previously reported total retinal area of 1,134 mm$^2$ [25] or 1,094 mm$^2$ [4]. However, if the empirical factor 1.5 is applied, the 180° visual-angle corresponds to 270°. Using the Equation 7, the 270° eye-angle corresponds to a retinal area of 1298 mm$^2$, which is ~15% overestimated compared to the reported 1,134 mm$^2$ [25] or 1,094 mm$^2$ [4]. Moreover, 288 µm per visual-angle has been well established to estimate the retinal dimension [4]. Using the conversion factor 1.5, the Equation 6 can perfectly match to 288 µm per visual-angle at the central field. However, for peripheral retina, As the conversion factor is gradually reduced up to 192 µm per degree at the far peripheral field (Fig. 5). This is consistent to the effect of the conversion factors on the eye-angle (Fig. 4) and retinal area (Fig. 6) estimations.

## 5. Conclusion

Reliable conversion between the visual-angle $\theta_v$ and eye-angle $\theta_e$ requires determined nodal point and spherical radius of the eye. Because each eye can be unique, it is difficult to have accurate information of the eye dimension and nodal point. For the consideration of generalization, traditional visual-angle can be a simple solution to specify the FOV of the fundus imager. However, eye-angle is required to estimate the imaged retinal distance and area. Using the model eye with average parameters, FOV conversion ratio $\theta_e/\theta_v$, angular conversion ratio $\Delta\theta_e/\Delta\theta_v$, and retinal conversion ratio ($\Delta d/\Delta\theta_v$) can be quantitatively evaluated. These conversion ratios are not linear over the visual field. The FOV conversion ratio $\theta_e/\theta_v$ can be changed from 1.51 in narrow field systems, to 1.34 in ultra-wide field imagers. Both angular conversion ($\Delta\theta_e/\Delta\theta_v$) and retinal conversion ($\Delta d/\Delta\theta_v$) ratio can have a 1.51 fold difference at the central and far peripheral regions. The empirical factor $\theta_e/\theta_v \approx 1.5$ is appropriate for imaging systems with visual-angle FOV < 80°. However, its use in ultra-wider field imagers may provide overestimated eye-angle FOV, and then retinal area.

## Funding

National Eye Institute (R01 EY023522, R01 EY029673, R01 EY030101, R01 EY030842, P30 EY001792); Research to Prevent Blindness; Richard and Loan Hill Endowment.

## Disclosures

The authors declare that there are no conflicts of interest related to this article.